\documentclass[authoryear,final,3p,twoside,onecolumn,times,a4paper]{elsarticle}
\usepackage[latin1]{inputenc}
\usepackage{xspace}
\usepackage[english]{babel}
\usepackage{amssymb,amsmath,amsfonts}
\usepackage{bm, bbm}
\usepackage{enumerate} 

\usepackage[hidelinks]{hyperref} \hypersetup{bookmarksnumbered, colorlinks=false,
  breaklinks, pdfborder=0 0 0, pdftitle=A new integral loss for Bayesian
  optimization, pdfauthor=Emmanuel Vazquez and Julien Bect, pdfkeywords=Global
  Optimization Gaussian process Expected Improvement}

\usepackage{notations}

\newcommand \CritName {$\mathrm{EI}^2$\xspace}

\newcommand \Fcal {\mathcal{F}}
\newcommand \xstar {x^{\star}}

\newcommand \loss {\varepsilon}

\usepackage{xcolor}

\begin{document}

\begin{frontmatter}

\title{A new integral loss function for Bayesian optimization}

\author{Emmanuel Vazquez and Julien Bect}

\makeatletter
\address{SUPELEC, 3 rue Joliot-Curie, 91192 Gif-sur-Yvette, France\\
email: \{firstname\}.\{lastname\}@\hskip 0.6pt supelec.fr}
\makeatother

\begin{abstract}
  We consider the problem of maximizing a real-valued continuous function $f$
  using a Bayesian approach. Since the early work of Jonas Mockus and Antanas
  \v{Z}ilinskas in the 70's, the problem of optimization is usually formulated
  by considering the loss function $\max f - M_n$ (where $M_n$ denotes the best
  function value observed after $n$ evaluations of~$f$). This loss function puts
  emphasis on the value of the maximum, at the expense of the location of the
  maximizer. In the special case of a one-step Bayes-optimal strategy, it
  leads to the classical Expected Improvement (EI) sampling
  criterion. This is a special case of a Stepwise Uncertainty Reduction (SUR)
  strategy, where the risk associated to a certain uncertainty measure (here,
  the expected loss) on the quantity of interest is minimized at each step of
  the algorithm. In this article, assuming that~$f$ is defined over a measure
  space~$\left( \XX, \lambda \right)$, we propose to consider instead the
  integral loss function $\int_{\XX} (f - M_n)_{+}\, d\lambda$, and we show that
  this leads, in the case of a Gaussian process prior, to a new numerically
  tractable sampling criterion that we call \CritName (for Expected Integrated
    Expected Improvement). A numerical experiment illustrates that
  a SUR strategy based on this new sampling criterion reduces the error on 
  both the value and the location of the maximizer faster than the EI-based strategy.
\end{abstract}


\end{frontmatter}

\section{Introduction}
\label{sec:introduction}

Let~$f: \XX \to \RR$ be a real-valued continuous function defined on
a compact subset~$\XX$ of~$\RR^d$, $d \geq 1$. We consider the problem of
finding an approximation of the maximum of $f$,
\begin{equation*}
  M = \max_{x\in\XX} f(x)\,,  
\end{equation*}
and of the set of maximizers,
\begin{equation*}
  \xstar\in \argmax_{x\in\XX} f(x)\,,
\end{equation*}
using a sequence of queries of the value of~$f$ at points $X_1,\, X_2,\, \ldots
\in \XX$. At iteration $n+1$, the choice of the evaluation point $X_{n+1}$ is
allowed to depend on the results~$f(X_1),\, \ldots,\, f(X_n)$ of the evaluation
of~$f$ at $X_1,\, \ldots,\, X_{n}$. Thus, the construction of an optimization
strategy $\Xv = \left( X_1,\, X_2,\, \ldots \right)$ can be seen as a sequential
decision problem.

We adopt the following Bayesian approach for constructing $\Xv$. The unknown
function~$f$ is considered as a sample path of a random process $\xi$ defined on
some probability space $(\Omega, \mathcal{B}, \P_0)$, with parameter $x \in
\XX$. For a given $f$, the efficiency of a strategy $\Xv$ can be measured in
different ways. For instance, a natural loss function for measuring the
performance of $\Xv$ at iteration $n$ is
\begin{equation}
  \label{eq:1}
   \loss_{n}(\Xv, f) = M - M_n\,,
\end{equation}
with $M_{n} = \max \left( f(X_1),\, \ldots,\, f(X_n) \right)$. The choice of a
loss function~$\loss_n$, together with a random process model, makes it possible
to define the following one-step Bayes-optimal strategy:
\begin{equation}
  \label{eq:one-step-lookahead}
  \left\{
    \begin{array}{l}
  X_1 = x_{\rm init}\\
  X_{n+1} \;=\;  \argmin_{x_{n+1} \in \XX}\, 
  \EE_n\, \Bigl(\loss_{n+1}(\Xv,\, \xi) \,\big|\, X_{n+1} = x_{n+1}
  \Bigr)\,,\quad \forall n\geq 1,\,      
    \end{array}\right.
\end{equation}
where $\EE_n$ denotes the conditional expectation with respect to the
$\sigma$-algebra $\FF_n$ generated by the random variables $X_1,\, \xi(X_1),\,
\ldots, X_n,\, \xi(X_n)$. This Bayesian decision-theoretic point of view has
been initiated during the 70's
by the work of Jonas Mockus and Antanas \v{Z}ilinskas \citep[see][and references
therein]{mockus:78:abmse, mockus:89:bagota}.

For instance, consider the loss defined by~\eqref{eq:1}. Then, at iteration $n +
1$, the strategy~\eqref{eq:one-step-lookahead} can be written as
\begin{eqnarray}  
  X_{n+1} &=&
  \argmin_{x_{n+1} \in \XX}\, \EE_n \left( M - M_{n+1} \mid X_{n+1} =
    x_{n+1}\right)
  \label{eq:from-Loss-to-EI} \\  
  &=& \argmax_{x_{n+1} \in \XX}\, \EI_n \left( x_{n+1} \right),
  \nonumber
\end{eqnarray}
where $\EI_n \left( x \right) \eqdef \EE_n \bigl( \max \left( \xi(x) - M_n,\, 0 \right) \bigr)$
is the \emph{Expected Improvement} (EI) criterion, introduced by
\cite{mockus:78:abmse} and later popularized through the EGO algorithm
\citep{jones:98:ego}, both in the case of Gaussian process models (for
which~$\EI_n \left( x \right)$ admits a closed-form expression as a function of the
posterior mean and variance of~$\xi$ at~$x$).

% Side remark: there exist Bayesian optimization strategies that do not
% explicit a definition of a loss function. For instance: P-algorithm, 
% GP-UCB\ldots 

The contribution of this paper is a new loss function for evaluating the
efficiency of an optimization strategy, from which we can derive, in the
case of a Gaussian process prior, a numerically tractable sampling
criterion for choosing the evaluations points according to a one-step
Bayes-optimal strategy. Section~\ref{sec:an-integral-loss} explains our
motivation for the introduction of a novel loss function, and then
proceeds to present the loss function itself and the associated sampling
criterion. The numerical implementation of this new sampling criterion
is discussed in Section~\ref{sec:analysis}. Finally,
Section~\ref{sec:illustr} presents a one-dimensional example that
illustrates qualitatively the effect of using our new loss function,
together with a numerical study that assesses the performance of the 
criterion from a statistical point of view on a set of sample paths of a
Gaussian process.

\section{An integral loss function}
\label{sec:an-integral-loss}

Observe that~\eqref{eq:from-Loss-to-EI} can be rewritten as
\begin{equation}
  \label{eq:EI-is-SUR}
  X_{n+1} \;=\;
  \argmin_{x_{n+1} \in \XX}\, \EE_n \left( H_{n+1} \mid X_{n+1} = x_{n+1} \right),
\end{equation}
with $H_n = \EE_n \left( M - M_n \right)$. The $\Fcal_{n+1}$-measurable random
variable $H_{n+1}$ in the right-hand side of~\eqref{eq:EI-is-SUR} can be seen as
a measure of the uncertainty about~$M$ at iteration~$n + 1$: indeed, according to
Markov's inequality, $M \in \left[ M_{n+1}; M_{n+1} + H_{n+1} / \delta \right]$
with probability at least $1 - \delta$ under~$\P_{n+1}$. Thus, this strategy is
actually a special case of \emph{stepwise uncertainty reduction}
\citep{villemonteix:2009:IAGO, bect:2012:stco, chevalier}.

In a global optimization problem, it is generally of interest to obtain a good
approximation of \emph{both~$M$ and $\xstar$}. The classical loss function
$\loss_n = M - M_n$ is not very satisfactory from this respect, since the
associated uncertainty measure~$H_n = \EE_n \left( M - M_n \right)$ puts all the
emphasis on~$M$, at the expense of~$\xstar$. Other uncertainty measures
have been proposed recently, which take the opposite approach and focus
on~$\xstar$ only \citep{villemonteix:2009:IAGO, picheny:2014:MO, picheny2014}.

Assume now that~$\XX$ is endowed with a finite positive measure~$\lambda$ (e.g.,
Lebesgue's measure restricted to~$\XX$), and let us remark that the
classical loss function~\eqref{eq:1} is proportional to
$\lambda(\XX)\, (M - M_n)$, that is, to the area of the
hatched region in Figure~1a. This illustrates that $H_n = \EE_n (\loss_n)$ is
only a coarse measure of the uncertainty about the pair $(M, \xstar)$. We
propose to use instead the integral loss function
\begin{equation}
  \label{eq:2}
  \loss'_n(\Xv, f) = \int_{\XX} (f(x) - M_n)_{+}\, \lambda(\dx),
\end{equation}
where $z_+ \eqdef \max \left(z, 0 \right)$. This new loss function is depicted in
Figure~1b. The associated uncertainty measure $H'_n = \EE_n \left( \loss'_n
\right)$ should, intuitively, provide a finer measure of the uncertainty about
the pair $(M, \xstar)$ and thereby lead to better optimization algorithms. The
corresponding stepwise uncertainty reduction strategy can be written as
\begin{eqnarray}
  X_{n+1} &=&  \argmin_{x_{n+1} \in \XX}\, \EE_n \left(
    \int_{\XX} (\xi(y) - M_{n+1})_{+}\, \lambda(\dy)
    \bmid X_{n+1} = x_{n+1} \right)\nonumber\\
  &=& \argmin_{x_{n+1} \in \XX}\, \EE_n \left(
    \,\int_{\XX}  \EE_{n+1}\bigl((\xi(y) - M_{n+1})_{+}\bigr)\, \lambda(\dy)
    \bmid  X_{n+1} = x_{n+1}
  \right)\nonumber \\
  &=& \argmin_{x_{n+1} \in \XX}\, \aleph_{n} \left( x_{n+1} \right),
  \label{eq:new-strategy}
\end{eqnarray}
where
\begin{equation}
  \aleph_n \left( x_{n+1} \right)
  \eqdef 
  \EE_n\left(
    \int_{\XX}  \EI_{n+1} \left( y \right)\, \lambda(\dy)  ~\Big|~
    X_{n+1} = x_{n+1} \right)\, \label{eq:def-aleph-1}
\end{equation}
is a new sampling criterion than we call \CritName (for Expected Integrated Expected
Improvement).
Note that the strategy~\eqref{eq:new-strategy} is very different in spirit from the classical one,
associated to the EI criterion. Indeed, while the classical strategy selects a
point where the \emph{current} EI is \emph{maximal}, the new strategy selects a
point where the integral of the \emph{future} EI is 
\emph{minimal}, in expectation.

\paragraph{Remark} The sampling criterion defined by~\eqref{eq:def-aleph-1} is a
one-point sampling criterion; that is, a sampling criterion for use in a fully
sequential setting. A multi-point sampling criterion can be defined similarly,
for use in a batch-sequential setting:
\begin{equation}
  \label{eq:batch-crit}
  \aleph_{n, r} \left( x_{n+1}, \ldots, x_{n+r} \right)
  \eqdef 
  \EE_n\left(
    \int_{\XX}  \EI_{n+r} \left( y \right)\, 
    \lambda(\dy) \bmid X_{n+1} = x_{n+1}, \ldots, X_{n+r} = x_{n+r}
  \right)
\end{equation}
(see \citet{chevalier:2013:qEI, chevalier} and references therein for more
information on multi-point stepwise uncertainty reduction strategies).

\begin{figure}[htbp]
  \hspace{-1cm}
  \begin{tabular}{cc}
    \psfrag{x}{\small $\XX$}
    \psfrag{M}{\small $M$}
    \psfrag{n}{\small $M_n$}

    \includegraphics[width=0.58\textwidth]{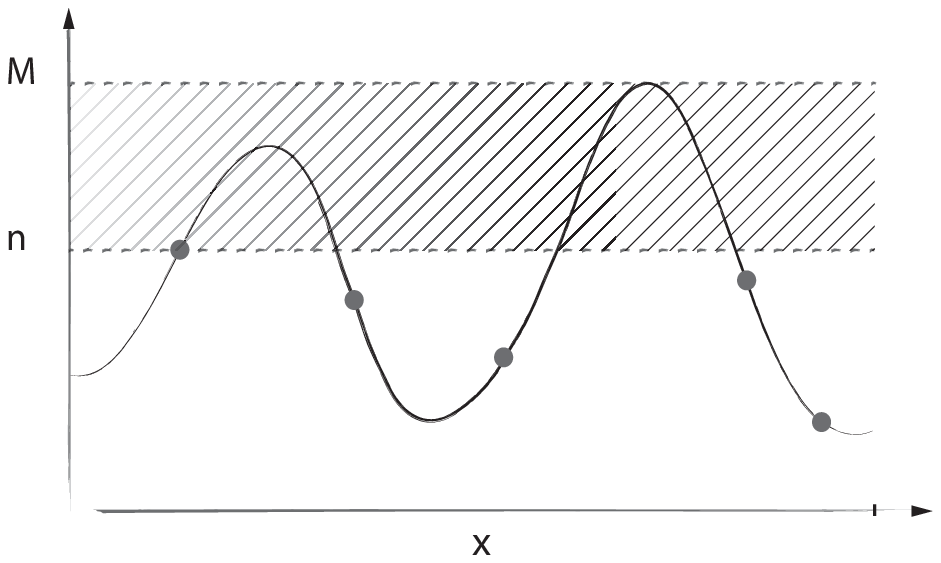}
    &
    \hspace{-1.2cm}\psfrag{x}{\small $\XX$}\psfrag{M}{\small
      $M$}\psfrag{n}{\small $M_n$}
    \includegraphics[width=0.58\textwidth]{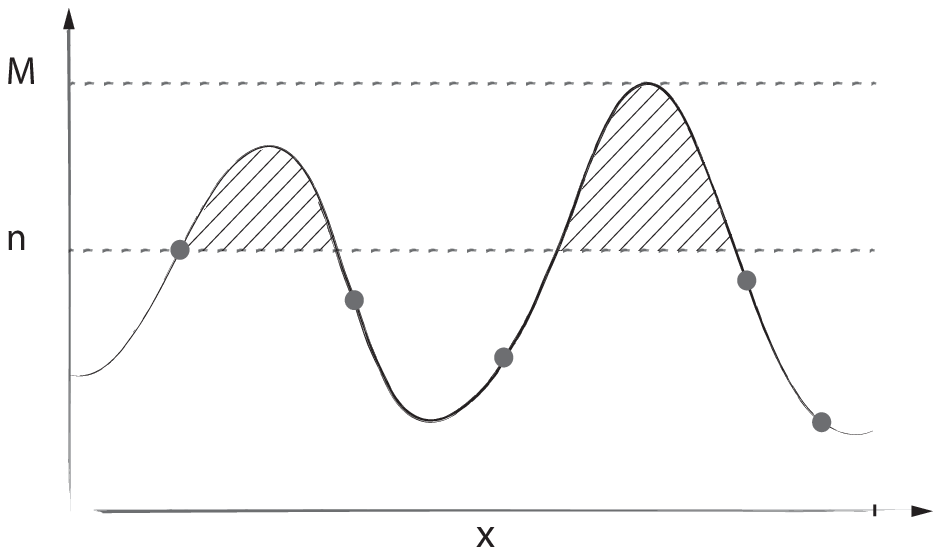}\\
    (a) & (b)
  \end{tabular}
  \caption{A diagrammatic interpretation of the loss functions
    $\loss_n$ (left plot) and $\loss_n^{\prime}$ (right plot).}
  \label{fig:1}
\end{figure}

\section{Numerical approximation of the sampling criterion}
\label{sec:analysis}

Numerical approximations of the sampling criterion $\aleph_n$ can be obtained
with an acceptable computational complexity when~$\xi$ is a Gaussian
process. Rewrite~\eqref{eq:def-aleph-1} as
\begin{equation}
  \label{eq:newcrit-intEEI}
  \aleph_n \left( x_{n+1} \right)
  = \int_{\XX}  \EEI_n \left( y; x_{n+1} \right)\, \lambda(\dy)\,,
\end{equation}
where $\EEI_n \left( y; x_{n+1} \right)$, which we shall call the \emph{Expected Expected
  Improvement} (EEI) at~$y \in \XX$ given a new evaluation  at~$x_{n+1}
\in \XX$, is defined by
\begin{equation}
  \label{eq:EEI}
  \EEI_n \left( y; x_{n+1} \right)
  \eqdef
  \EE_n \left(\,
    \EI_{n+1} \left( y \right)
    \bmid X_{n+1} = x_{n+1}
  \right)\,.
\end{equation}
(Note that $\EEI_n \left( y; x_{n+1} \right) \neq \EI_n (y)$ because of
the implicit dependency of~$\EI_{n+1} \left( y \right)$ on the future
maximum~$M_{n+1}$.)

It turns out that~$\EEI_n \left( y; x_{n+1} \right)$ can be expressed in closed
form, as a function of the posterior mean and covariance of~$\xi$, using the
special functions~$\Phi$, the cumulative distribution function of the univariate
standard normal distribution, and~$\Phi_2$, the cumulative distribution function of
the bivariate standard normal distribution. To see this, observe that
\begin{equation}
  \left( \xi(y) - M_{n+1} \right)_+ 
  =
  \widetilde M_{n+2} - M_{n+1}
  =
  \bigl(\widetilde M_{n+2} - M_n \bigr)
  - \bigl( M_{n+1} - M_n \bigr),
\end{equation}
where $\widetilde M_{n+2} = \max \bigl( M_{n+1},\, \xi(y) \bigr)$. Therefore, we have
\begin{equation}
  \label{eq:EEI-id-rem}
  \EEI_n \left( y; x_{n+1} \right) = \EE_n \bigl( (\xi(y) -
    M_{n+1})_{+} \mid X_{n+1} = x_{n+1}\bigr) 
  = 
  \EI_{n,\,2} \left( x_{n+1}, y \right) - \EI_n \left( x_{n+1} \right),
\end{equation}
where $\EI_{n,\,r}$ denotes the $r$-point expected improvement criterion:
\begin{equation}
  \label{eq:multi-point-EI}
  \EI_{n,\,r} \left( x_{n+1}, \ldots, x_{n+r} \right)
  \eqdef
  \EE_n \left( 
    M_{n+r} - M_n \bmid 
    X_{n+k} = x_{n+k},\, 1 \le k \le r
  \right).
\end{equation}
Equation~\eqref{eq:EEI-id-rem} makes it possible to compute $\EEI_n \left( y;
  x_{n+1} \right)$ using the closed-form expression obtained for the multi-point~EI by \citet{chevalier:2013:qEI}.

Assuming that $\lambda(\XX) < +\infty$, a simple idea for the computation of the
integral over $\XX$ in~\eqref{eq:newcrit-intEEI} is to use a Monte Carlo approximation:
\begin{equation*}
  \aleph_{n} \left( x_{n+1} \right) \;\approx\; 
  \frac{\lambda(\XX)}{m}\; \sum_{i=1}^m\; 
  \EEI_n \left( Y_i; x_{n+1} \right)
\end{equation*}
where $(Y_i)_{1 \le i \le m}$ is a sequence of independent random variables
distributed according to~$\lambda\left( \cdot \right) / \lambda(\XX)$.
Since $\aleph_n$ has also to be minimized over $\XX$, we can also use the sample $(Y_i)_{1 \le i
  \le m}$ to carry out a simple stochastic optimization. In practice however, we
would recommend to use a more advanced sequential Monte Carlo method, in the
spirit of that described in~\cite{benassi2012bayesian}
and~\citet{benassi2013nouvel}, to carry out both the integration and the
optimization steps.

\paragraph{Remark}
Equations~\eqref{eq:newcrit-intEEI}--\eqref{eq:multi-point-EI} are easily
generalized to batch sequential optimization. Define a multi-point EEI by
\begin{equation*}
  \EEI_{n,\,r} \left( y; x_{n+1}, \ldots, x_{n+r} \right)
  \eqdef
  \EE_n \left(\,
    \EI_{n+r} \left( y \right)
    \bmid X_{n+k} = x_{n+k},\, 1 \le k \le r
  \right)\,.
\end{equation*}
We have
\begin{equation*}
  \EEI_{n,\,r} \left( y; x_{n+1}, \ldots, x_{n+r} \right)\,
  = \EI_{n,\,r+1} \left( x_{n+1}, \ldots, x_{n+r}, y \right) 
  - \EI_n \left( x_{n+1}, \ldots, x_{n+r} \right)\,.
\end{equation*}
Then, we can express a multi-point version of the sampling
criterion~(\ref{eq:def-aleph-1}) as
\begin{equation*}
\label{eq:batch-crit}
  \aleph_{n,\,r} \left( x_{n+1}, \ldots, x_{n+r} \right)
  =
  \int_\XX \EEI_{n,\,r} \left( y; x_{n+1}, \ldots, x_{n+r} \right)\,
  \lambda\left( \dy \right)\,.
\end{equation*}

\section{Numerical study}
\label{sec:illustr}

The numerical results presented in this section have been obtained with STK
\citep{stk}, a free GPL-licenced Matlab/Octave kriging toolbox.

\begin{figure}[tbp]
  \centering
  \begin{tabular}{cc}
    \psfrag{x}[c]{\small$x$}
    \psfrag{xi}[t][t]{\small$\xi(x)$}
    \psfrag{rho}[t][t]{\small$-\bigl(M_n + \EI_n\bigr)$}
    \includegraphics[width=0.49\textwidth]{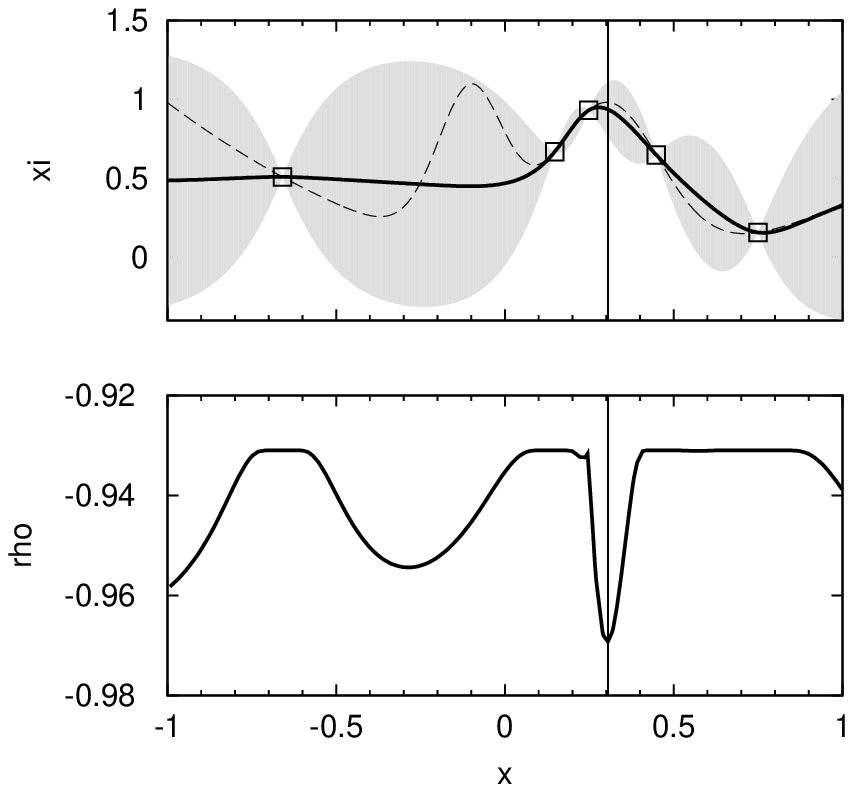} &
    \psfrag{x}[c]{\small$x$}
    \psfrag{xi}[t][t]{\small$\xi(x)$}
    \psfrag{rho}[t][t]{\small$\aleph_{n}(x)$}
    \includegraphics[width=0.49\textwidth]{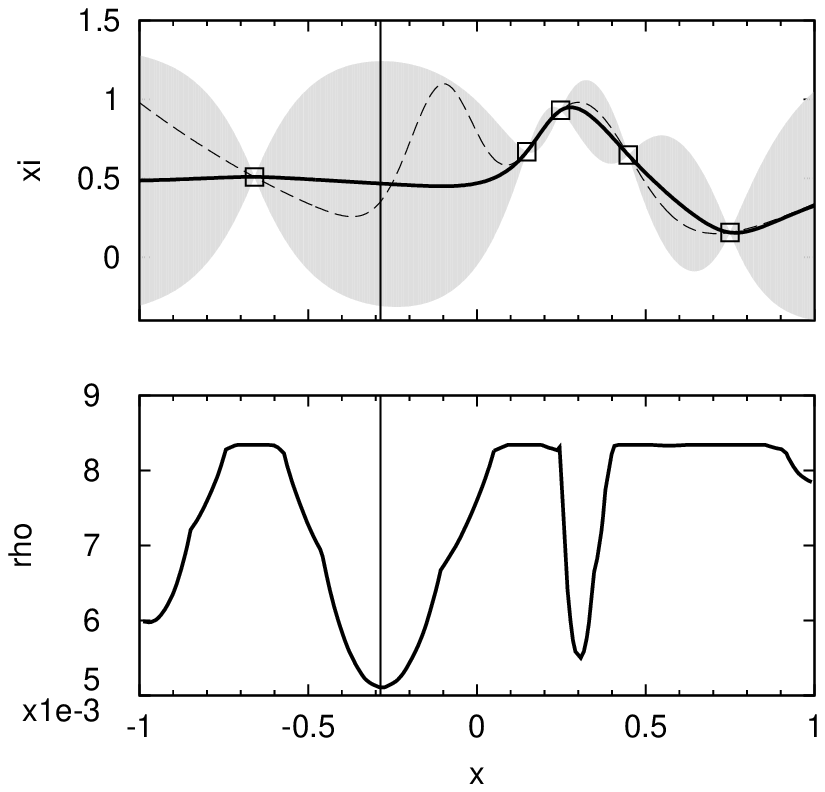}
  \end{tabular}
  \caption{Assessment of the behavior of the sampling criterion~$\aleph_n$
    (\emph{bottom, right}) against that of~$\EI_n$ (\emph{bottom, left}). The
    objective is to maximize the function $f:x\in [-1,1]\mapsto
    \bigl(0.8x-0.2\bigr)^2+ \exp\bigl(-\frac{1}{2} \abs{x +
      0.1}^{1.95}/0.1^{1.95}\bigr)+\exp\bigl(-\frac{1}{2}(2x-0.6)^2/0.1\bigr) -
    0.02$ (\emph{top}, dashed line). Evaluations points are represented by
    squares; the posterior mean $\xihat_{n}$ is represented by a solid  line;
    95\% credible intervals computed using $s_n$ are represented by gray
    areas. The next evaluation point will be chosen at the minimum of the
    sampling criterion (vertical solid line).}
  \label{fig:2}
\end{figure}

First, we present a simple one-dimensional illustration, whose aim is to
contrast qualitatively the behaviour of a sampling strategy based on the
\CritName criterion~$\aleph_n$ with that of the classical EI-based strategy.
Figure~\ref{fig:2} depicts a situation where
there is a large expected improvement in a small region of the search
domain, and a smaller expected improvement over a large
region of the search domain. In such a situation, the new sampling
criterion~$\aleph_n$ favors the large region with a smaller expected
improvement, thereby inducing a better exploration of the search domain
than~$\EI_n$.

\begin{figure}[tbp]
  \centering
  \begin{tabular}{cc}
  \psfrag{iter}[bc]{$n$}
  \psfrag{err}[bc]{$\ns{x^{\star} - x_{n}^{\star}}$}
  \includegraphics[width=0.5\textwidth]{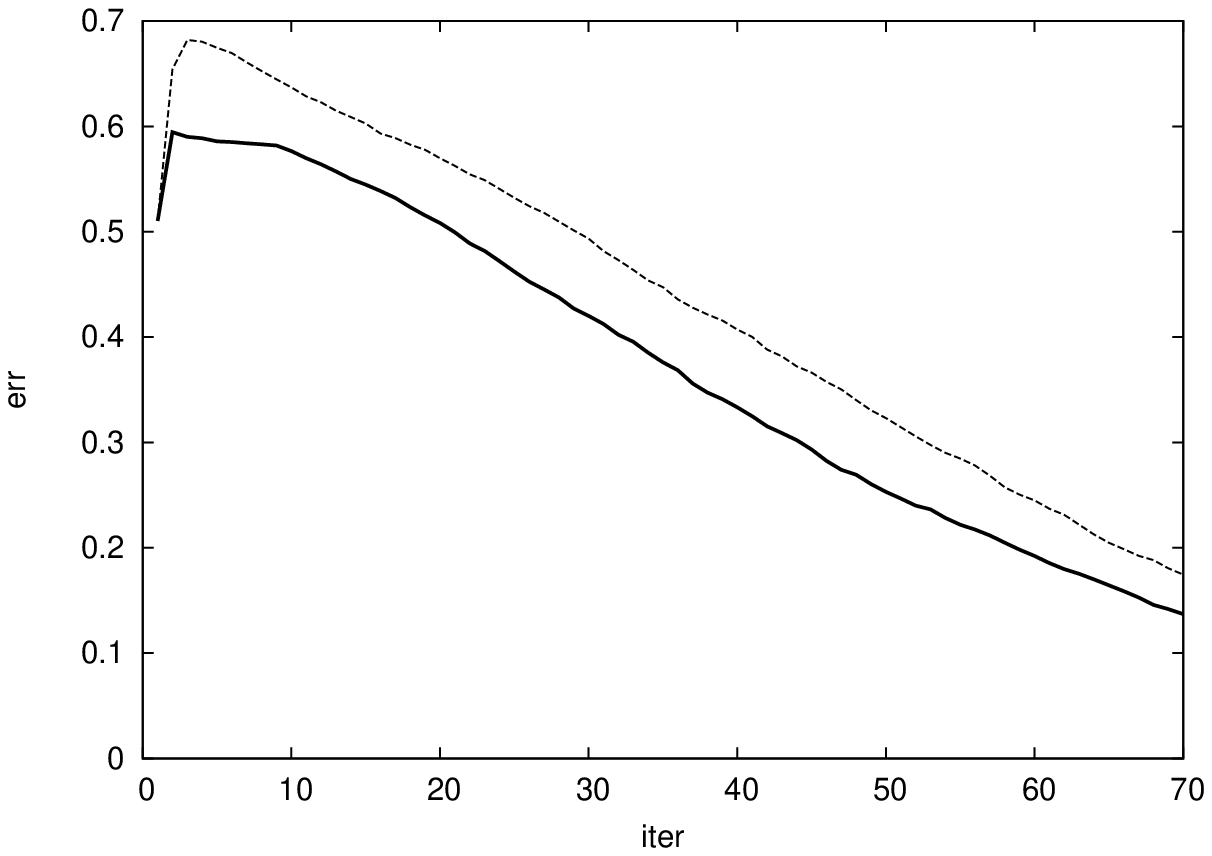} & 
  \psfrag{iter}[bc]{$n$}
  \psfrag{err}[bc]{$M - M_n$}
  \includegraphics[width=0.5\textwidth]{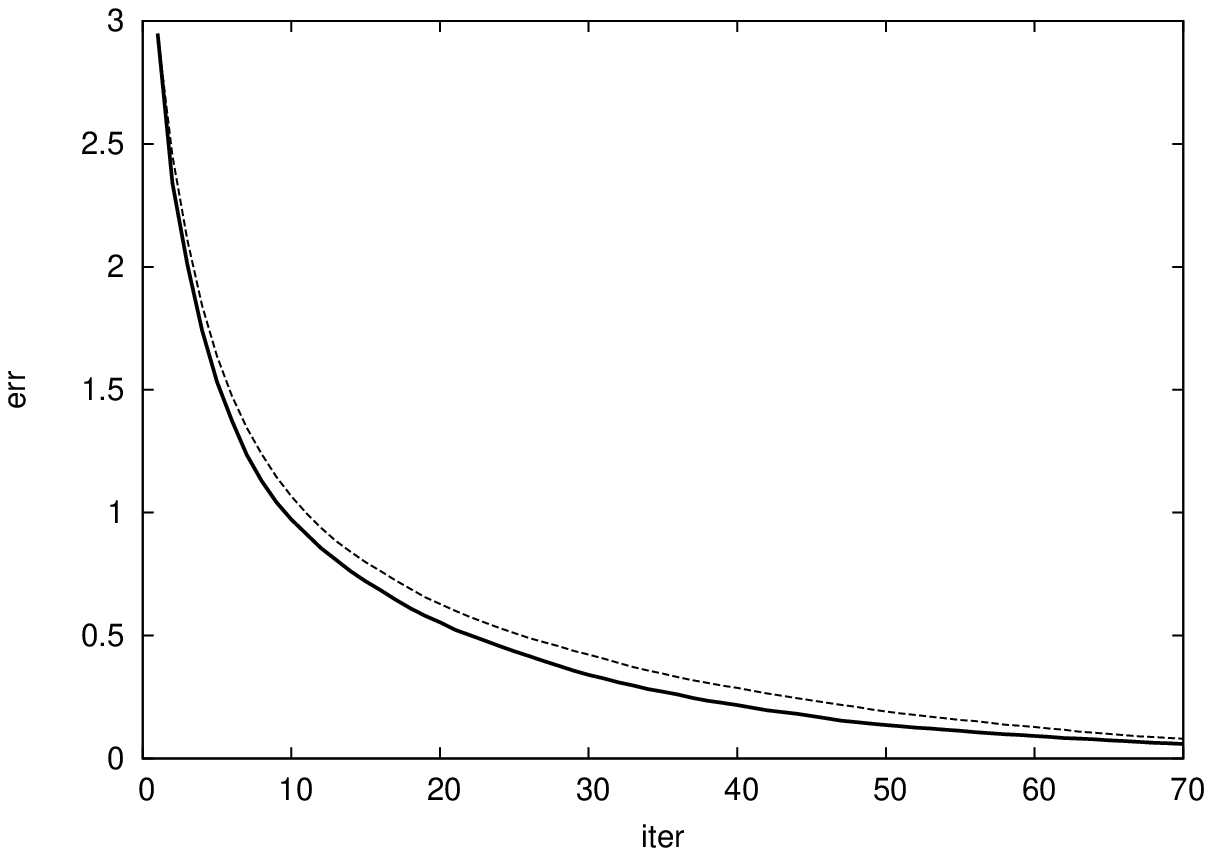}    
  \end{tabular}

  \caption{Approximation errors of $x^{\star}$ (\emph{left}) and $M$
    (\emph{right}) using the sampling criteria~$\aleph_n$ (solid line)
    and~$\EI_n$ (dashed line), as a function of the number of
    evaluations $n$. More precisely, each plot represents an average
    approximation error obtained on a testbed of $2700$ sample paths of
    a Gaussian process on $\RR^3$, with zero-mean and isotropic Mat\'ern
    covariance function, simulated on a set of $1000$ points in
    $[0,1]^3$.}
  \label{fig:3}
\end{figure}

Figure~\ref{fig:3} represents, for both strategies, the average
approximation error obtained on a testbed of $2700$ sample paths of a
Gaussian process on $\RR^d$, $d=3$, with zero-mean and isotropic
Mat\'ern covariance function, simulated on a set of $m=1000$ points in
$[0,1]^d$. The isotropic form of the Mat\'ern covariance on $\RR^{d}$
may be written as $k(x,y) = \sigma^2 r_\nu(\ns{x - y}/\beta)$, with
$r_{\nu}: \RR^{+} \to \RR^{+}$ such that, $\forall h \geq 0$,
$$
r_{\nu}(h) =
\frac{1}{2^{\nu-1}\Gamma(\nu)}\left(2\nu^{1/2}h\right)^\nu
\mathcal{K}_\nu\left(2\nu^{1/2} h \right)\,,
$$
where $\Gamma$ is the Gamma function and $\mathcal{K}_\nu$ is the
modified Bessel function of the second kind of order~$\nu$. Here,
$\sigma^2 = 1.0$, $\beta = (4\!\cdot\!10^{-2} \Gamma(d/2+1) / \pi^{d/2}
)^{1/d} \approx 0.2$ and $\nu=6.5$.  For each optimization strategy, we
use the same covariance function for $\xi$ than that used to generate
the sample paths in the testbed. Before running the optimization
strategies, an initial evaluation point $x_1$ is set at the center of
$[0,1]^{d}$.  For each sample path $f$, and each $n\geq 1$, the
estimator $x_n^{\star}$ of $x^{\star}$ is defined as $x_n^{\star} =
\argmax_{x \in \{x_1, \ldots, x_n \}} f(x)$. Thus, $\ns{x^{\star} -
  x_{n}^{\star}}$ is not a decreasing function of $n$ in
general. Figure~\ref{fig:3} shows that the approximation errors $M -
M_n$ and $\ns{x^{\star} - x_{n}^{\star}}$ decrease approximately at the
same rate for both strategies; however, the Euclidean distance of
$x_n^{\star}$ to $x^{\star}$ is significantly smaller in the case of the
new strategy.

\section*{References}
\bibliographystyle{elsarticle-harv}
\bibliography{refs}

\begin{thebibliography}{12}
\expandafter\ifx\csname natexlab\endcsname\relax\def\natexlab#1{#1}\fi
\expandafter\ifx\csname url\endcsname\relax
  \def\url#1{\texttt{#1}}\fi
\expandafter\ifx\csname urlprefix\endcsname\relax\def\urlprefix{URL }\fi

\bibitem[{Bect et~al.(2012)Bect, Ginsbourger, Li, Picheny, and
  Vazquez}]{bect:2012:stco}
Bect, J., Ginsbourger, D., Li, L., Picheny, V., Vazquez, E., 2012. Sequential
  design of computer experiments for the estimation of a probability of
  failure. Statistics and Computing 22~(3), 773--793.

\bibitem[{Bect et~al.(2014)Bect, Vazquez, et~al.}]{stk}
Bect, J., Vazquez, E., et~al., 2014. {STK}: a {S}mall ({M}atlab/{O}ctave)
  {T}oolbox for {K}riging. {R}elease 2.1.
\newline\urlprefix\url{http://kriging.sourceforge.net}

\bibitem[{Benassi(2013)}]{benassi2013nouvel}
Benassi, R., 2013. Nouvel algorithme d'optimisation bay{\'e}sien utilisant une
  approche monte-carlo s{\'e}quentielle. Ph.D. thesis, Sup{\'e}lec.

\bibitem[{Benassi et~al.(2012)Benassi, Bect, and Vazquez}]{benassi2012bayesian}
Benassi, R., Bect, J., Vazquez, E., 2012. Bayesian optimization using
  sequential {M}onte {C}arlo. In: Learning and Intelligent Optimization. 6th
  International Conference, LION 6, Paris, France, January 16-20, 2012, Revised
  Selected Papers. Vol. 7219 of Lecture Notes in Computer Science. Springer,
  pp. 339--342.

\bibitem[{Chevalier et~al.(2013)Chevalier, Bect, Ginsbourger, Vazquez, Picheny,
  and Richet}]{chevalier}
Chevalier, C., Bect, J., Ginsbourger, D., Vazquez, E., Picheny, V., Richet, Y.,
  2013. Fast parallel kriging-based stepwise uncertainty reduction with
  application to the identification of an excursion set. Technometrics, 22
  pages, accepted for publication, posted online: 21 Nov 2013.

\bibitem[{Chevalier and Ginsbourger(2013)}]{chevalier:2013:qEI}
Chevalier, C., Ginsbourger, D., 2013. Fast computation of the multi-points
  expected improvement with applications in batch selection. In: Nicosia, G.,
  Pardalos, P. (Eds.), Learning and Intelligent Optimization. LNCS. Springer,
  pp. 59--69.

\bibitem[{Jones et~al.(1998)Jones, Schonlau, and Welch}]{jones:98:ego}
Jones, D.~R., Schonlau, M., Welch, W.~J., 1998. Efficient global optimization
  of expensive black-box functions. J. Global Optim. 13, 455--492.

\bibitem[{Mockus(1989)}]{mockus:89:bagota}
Mockus, J., 1989. Bayesian approach to Global Optimization: Theory and
  Applications. Kluwer Acad. Publ., Dordrecht-Boston-London.

\bibitem[{Mockus et~al.(1978)Mockus, Tiesis, and
  \v{Z}ilinskas}]{mockus:78:abmse}
Mockus, J., Tiesis, V., \v{Z}ilinskas, A., 1978. The application of {B}ayesian
  methods for seeking the extremum. In: Dixon, L., Szego, G. (Eds.), Towards
  Global Optimization. Vol.~2. North Holland, New York, pp. 117--129.

\bibitem[{Picheny(2014{\natexlab{a}})}]{picheny:2014:MO}
Picheny, V., 2014{\natexlab{a}}. Multiobjective optimization using {G}aussian
  process emulators via stepwise uncertainty reduction. arXiv:1310.0732 (to
  appear in \emph{Statistics and Computing}).

\bibitem[{Picheny(2014{\natexlab{b}})}]{picheny2014}
Picheny, V., 2014{\natexlab{b}}. A stepwise uncertainty reduction approach to
  constrained global optimization. In: Proceedings of the 17th International
  Conference on Artificial Intelligence and Statistics (AISTATS), 2014,
  Reykjavik, Iceland. Vol.~33. JMLR: W\&CP, pp. 787--795.

\bibitem[{Villemonteix et~al.(2009)Villemonteix, Vazquez, and
  Walter}]{villemonteix:2009:IAGO}
Villemonteix, J., Vazquez, E., Walter, E., 2009. An informational approach to
  the global optimization of expensive-to-evaluate functions. Journal of Global
  Optimization 44~(4), 509--534.

\end{thebibliography}

\end{document}